\documentclass[pra,twocolumn]{revtex4-1}

\usepackage[english]{babel}
\usepackage[utf8]{inputenc}
\usepackage{graphicx}
\usepackage{graphics}
\usepackage{braket}
\usepackage{graphicx}
\usepackage[parfill]{parskip}
\usepackage{array}
\usepackage{fancyhdr}
\usepackage{fullpage}
\usepackage{braket}
\usepackage{tabulary}
\usepackage{setspace}
\usepackage{bbold}
\usepackage{amsmath}

\begin{document}
\title{Non-ideal teleportation of tripartite entanglement: Einstein-Podolsky-Rosen {\it versus} Greenberger-Horne-Zeilinger schemes}
\author{Márcio M. Cunha, E. A. Fonseca, and Fernando Parisio}
\email[]{parisio@df.ufpe.br}
\affiliation{Departamento de F\'{\i}sica, CCEN,  Universidade Federal 
de Pernambuco, 50670-901 , Recife, PE, Brazil}

\begin{abstract}
Channels composed by Einstein-Podolsky-Rosen (EPR) pairs are capable of teleporting arbitrary multipartite states. The question arises whether EPR channels are also optimal against imperfections. In particular, the teleportation of Greenberger-Horne-Zeilinger states (GHZ) requires three EPR states as the channel and full measurements in the Bell basis. We show that, by using two GHZ states as the channel, it is possible to transport any unknown three-qubit state of the form $c_0|000\rangle+c_1|111\rangle$. The teleportation is made through measurements in the GHZ basis, and, to obtain deterministic results, in most of the investigated scenarios, four out of the eight elements of the basis need to be unambiguously distinguished. Most importantly, we show that when both, systematic errors and noise are considered, the fidelity of the teleportation protocol is higher when a GHZ channel is used in comparison to that of a channel composed by EPR pairs.

\end{abstract}
\maketitle

\section{Introduction}
Some findings have such a high degree of simplicity and relevance that become paradigms overnight. This is certainly the case of the quantum teleportation of an arbitrary, unknown qubit for the field of quantum information \cite{bennett,bowmeester,pirandola}. Of course, between the proof of possibility in principle \cite{bennett} and an actual teleportation in the laboratory \cite{bowmeester}, there are several fundamental and practical dificulties. For example, in the original scheme, full measurements in a maximally entangled basis are required to obtain success in every run. This task, however, cannot be executed via linear one-qubit elements only, e. g., phase shifters and beam splitters \cite{mattle, pan}. Because of this, the first experimental implementation was conditional, requiring postselection \cite{bowmeester} (for unconditional implementations see, e. g., \cite{barrett,riebe}). Even when complete measurements can be carried out, the channels and the measurement basis always present, to some extent, systematic (deviations from maximal entanglement) and random (environmental) imperfections. 

A way to circumvent the difficulties related to the presence of noise is to resort to redundancy. In classical communications it is usual to encode a single bit, say 0, as 000, so that error correction codes can recover the message with a prescribed success rate (note that an odd number of bits is required to avoid undecidable situations). Analogously, it is safer to encode the information contained in a single qubit $c_0|0\rangle+c_1|1\rangle$ in the larger state $c_0|000\rangle+c_1|111\rangle$ \cite{barnett}. In fact, more than three qubits would be necessary to enable the correction of an arbitrary single-qubit error \cite{barnett,steane}. 

Another important use of Greenberger-Horne-Zeilinger (GHZ) states is as heralded Einstein-Podolsky-Rosen (EPR) pairs, since $|000\rangle+|111\rangle=|+\rangle|\Phi^+\rangle+|-\rangle|\Phi^-\rangle$, with $|\pm\rangle=(|0\rangle\pm |1\rangle)/\sqrt{2}$ and $|\Phi^{\pm}\rangle=(|00\rangle\pm |11\rangle)/\sqrt{2}$, so that a $+ (-)$ detection in one of the parties heralds
the existence of EPR entanglement $|\Phi^{+}\rangle (|\Phi^{-}\rangle)$ between the other two parties.
In addition, several tasks in quantum computation demand more complex forms of entanglement, as for example, one-way quantum computing \cite{raussendorf}, for which graph states are needed.

It is, therefore, of evident interest to investigate efficient ways to teleport multiqubit entangled states. The most usual procedure is to employ as the channel a sufficient number of pairs of maximally entangled qubits, that is, EPR states. This is justifiable since bipartite entanglement is easier to prepare and serve as a universal resource for the teleportation of arbitrary states \cite{min}.
Particularly, the teleportation of three-particle states has been shown to be possible in several ways, being deterministic, for an ideal GHZ state via ideal EPR pairs \cite{Yang}, and, probabilistic for arbitrary three-particle states through a channel composed by non-maximally entangled EPR-like pairs \cite{fang}.
The general problem of teleporting $n$-partite states with $n$ EPR pairs has been addressed in \cite{ikram,min} and carried out experimentally for the case of two qubits in \cite{zhang}.

There are, however, works where instead of EPR pairs, more complex states are used as quantum channels. One qubit teleportation protocols are presented in \cite{moussa} and \cite{karlsson}, where GHZ states are employed, while in \cite{liang} a maximally entangled six-qubit state is used to teleport an arbitrary three-qubit state (see \cite{nie} for a scheme of controlled teleportation). In all these cases, the swapping operations that materialize the teleportation correspond to Bell measurements, that is, measurements in a bipartite entangled basis.

The question arises on what is the effect of systematically employing a channel of tripartite states and tripartite swapping operations in the teleportation of an unknown GHZ-like state. What are the final fidelities in comparison to a protocol that uses EPR  pairs and measurements?
In this work we survey on the transport of GHZ-like states, $c_0|000\rangle+c_1|111\rangle$, with $c_0$ and $c_1$ unknown. 

As a starting point, we develop a compact notation that encompass quite general protocols. We address the
teleportation of three partite states through two different types of channels: (i) three EPR-type states and (ii) two GHZ-type states (in a particular geometric configuration). 
In addition, in scenario (i) the swapping is done using three Bell measurements, while in scenario (ii) two GHZ measurements are required. 
We refer to the first scheme as ``3-EPR'' and to the second one as ``2-GHZ''. We show that in a totally ideal scenario 
(perfect channel and measurements without noise) the 3-EPR and 2-GHZ scheme lead to a fidelity of $100 \%$, although the latter does not require complete GHZ measurements. 
We then proceed to investigate the individual and combined effects of systematic deviations from maximal entanglement, both, in the channels and in the measurements, and of common types of noise. 

Throughout this work there will be nine qubits involved in each teleportation event. Six of these qubits (labeled 1 to 6) are close together, while the remaining three qubits (7, 8, and 9) are in a distant location. 
The GHZ-like state to be teleported is encoded in the odd-labeled qubits 1-3-5 and may be written as:
\begin{equation}
\label{input}
\ket{\phi}=\sum_{j=0,1}c_{j} \ket{jjj}.
\end{equation}
Since this state is spatially localized and no part of it need to be physically moved in a preliminary stage, we will assume hereafter that it is protected from 
noise. For the other six qubits, forming the quantum channel, the entanglement is distributed among distant parties. In the final part of this work, we consider that the 
channels are not perfectly isolated from the environment, at least in some preparatory stage. In this weak noise regime, where the probability of an error occurring in one
of the qubits is small, we find that the 2-GHZ scheme has a better performance.  It is worth mentioning that, only recently a comprehensive account of the effects of 
practically relevant environmental disturbances was provided in the simplest case of the teleport of a qubit through an EPR channel \cite{rigolim}.
\section{Quantum teleportation}
Let us briefly revisit the standard teleportation protocol to set our notation. It involves two distant parties, Alice and Bob, sharing a quantum channel consisting of a pair of entangled 
qubits in the state $\hat{\rho}_{ch}$. Alice intends to send an unknown quantum state $\ket{\phi}$ to Bob. She carries out a joint measurement in an orthonormal 
basis $\{\ket{\Phi_K}\}$ on her part of the channel and on the qubit  to be sent. She classically informs Bob about her outcome. Finally, Bob applies a local unitary operation $\hat{U}_K$ 
on his qubit. After each run, up to normalization, the state of Bob's qubit is given by:
\begin{equation}
\hat{\rho}_K= \hat{U}_K\mathrm{Tr}_{A}\left\{\left(\hat{P}_K\otimes \hat{1}_B\right)\ket{\phi}\bra{\phi}\otimes\hat{\rho}_{ch}\right\}\hat{U}_K^{\dagger},
\end{equation} 
where $\hat{P}_K$ is the projector $\ket{\Phi_K}\bra{\Phi_K}$ and $\mathrm{Tr}_{A}$ denotes the partial trace over Alice's system. 
Since, usually, the deviations from ideality are unknown, the unitary transformations $\hat{U}_K$ refer to the ideal case of 
maximally entangled channels and measurements.
The fidelity of the teleported state with respect to $\ket{\phi}$ reads 
\begin{equation}
F=\sum_K\mathrm{Tr}\left\{\ket{\phi}\bra{\phi}\hat{\rho}_K\right\}.
\end{equation}

In general $F$ depends on the parameters that characterize the input state, thus 
in order to get a state independent figure of merit, we uniformly average over all possible input states:
\begin{equation}
\label{AvFid}
\langle F\rangle=\frac{1}{V}\int dV F,
\end{equation}
here, $dV$ is the volume element on the space of quantum states and $V$ is the total volume.
\section{Fidelity of Three-Partite entanglement teleportation}
In this section we present details of the teleportation protocols and general fidelity expressions for both schemes proposed in this work: 3-EPR and 2-GHZ.
\subsection{3-EPR Scheme}
\begin{figure}
	\includegraphics[scale=0.45]{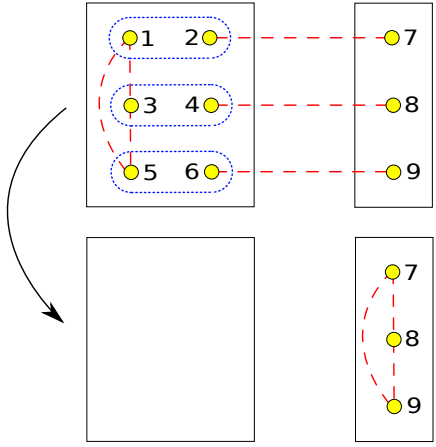}
	\caption{(color online) Teleportation of an unknown state $\sum_{j=0,1}c_{j} \ket{jjj}_{135}$ through a channel composed by three EPR states. The dotted ovals represent 
		complete measurements in the EPR basis $\{\ket{\Psi^{\mu}_{\lambda}}\}$ and the dashed lines, entanglement between the involved qubits.}
	\label{fig1}
\end{figure}

We consider that each of the qubits 2, 4, and 6 is a half of an entangled pair, the other parties being qubits 7, 8, and 9, respectively as summarized in figure \ref{fig1}.
The density operator of each pair of the channel is represented by:
\begin{equation}
\label{gamma1}
\hat{\rho}_{ab}=\sum_{klmn=0,1} \gamma_{klmn}^{(ab)}\ket{kl}\bra{mn},
\end{equation}
where the coefficient $\gamma_{klmn}^{(ab)}$ already includes the information on deviations from ideality.

The measurement basis is not assumed to be ideal in the sense that its four kets may not be maximally entangled, although always orthonormal. 
In this case, we use the EPR-like basis $\{\ket{\Psi^{\mu}_{\lambda}}\}$ (see the appendix) to represent the measurements on pairs of qubits. Any element of this 
basis may be expressed as: $\ket{\Psi_{\lambda}^{\mu}}=\sum_{j=0,1}(-1)^{\mu j}b_{\mu\oplus j}\ket{j,j\oplus\lambda}$, where $b_0=\cos\phi$, 
$b_1=\sin\phi$, and $\oplus$ stands for sum modulo 2. Note that for $\phi=\pi/4$, it corresponds to the Bell basis. Taking into account the three required measurements, the projector $\hat{P}_K$ is
\begin{equation}
\hat{P}_{K}= \ket{\Psi^{\mu}_{\lambda}}\bra{\Psi^{\mu}_{\lambda}}_{12}\otimes
\ket{\Psi^{\nu}_{\omega}}\bra{\Psi^{\nu}_{\omega}}_{34}\otimes
\ket{\Psi^{\epsilon}_{\tau}}\bra{\Psi^{\epsilon}_{\tau}}_{56}.
\end{equation}
For each output, after the appropriate exchange of classical information, local unitary transformations on the qubits 7, 8 and 9 
must be executed. The necessary transformations can be compactly expressed as:
\begin{equation}
\nonumber
\hat{U}_K= \sum_{klm=0,1}a_{\mu}^k a_{\nu}^l a_{\epsilon}^m\ket{k,l,m}\bra{k \oplus \lambda, l\oplus \omega, m \oplus \tau},
\end{equation}
with the coefficients $a_{\mu}^k$ defined in such a way that 
\begin{equation}
\nonumber
a_{\mu_1\mu_2\cdots \mu_m}^{j_1j_2\cdots j_n}\equiv(-1)^{(j_1+j_2+\cdots +j_n)(\mu_1 + \mu_2+\cdots+\mu_m)}.
\end{equation} 
The local character of the unitaries is made explicit via the equivalent, explicitly separable expression 
\begin{equation}
\hat{U}_{K}=\hat{\sigma}_z^{\mu}\hat{\sigma}_x^{\lambda}\otimes \hat{\sigma}_z^{\nu}\hat{\sigma}_x^{\omega} \otimes \hat{\sigma}_z^{\epsilon} \hat{\sigma}_x^{\tau},
\end{equation} 
where the standard notation for the Pauli matrices has been employed.
After some algebra, the fidelity reads:
\begin{widetext}
\begin{equation}
\label{F_EPR}
F=\sum_{\substack{klmn\mu\nu \\ \epsilon\lambda\omega\tau=0,1}}c_kc_l^*c_nc_m^* a_{\mu\nu\epsilon}^{klmn}\prod_{j=1}^3 b_{ \mu_j \oplus k}^* b_{\mu_j \oplus l}\gamma_{k\oplus \lambda_j,m\oplus\lambda_j,l\oplus\lambda_j,n\oplus\lambda_j}^{(a_jb_j)},
\end{equation}
\end{widetext}
where the index $j$ is related to the channel qubits and measurements, thus in order to get a compact expression, we wrote: $\{\lambda_1,\lambda_2,\lambda_3 \}=\{\lambda,\omega,\tau\}$ and $\{\mu_1,\mu_2,\mu_3\}=\{\mu,\nu,\epsilon\}$.

This expression will be used later to calculate fidelity of the 3-EPR teleportation protocol for several specific cases.

\subsection{2-GHZ Scheme}
In what follows we closely follow the previous procedure, this time, replacing the 3-EPR with the 2-GHZ scheme. 
The properties of this kind of tripartite state working as a part of a channel have been investigated in 
different contexts. GHZ channels have been shown to be capable of transporting a single qubit \cite{moussa, karlsson}, and, in reference \cite{ghosh}, e. g., 
it is shown that a bipartite entangled state, shared by two distant parties, $A$ and $C$ can be fully transported to other two distant parties, $B1$ and $B2$, whenever 
$A$, $B1$ and $B2$ share a GHZ state. Here, as in the previous section, we deal with the teleport of a GHZ-like state.

\begin{figure}[hpt]
\centering
\includegraphics[scale=0.45]{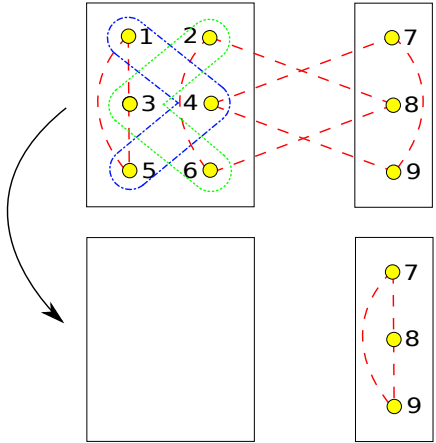}
\caption{(color online) Teleportation of an unknown state $\sum_{j=0,1}c_{j} \ket{jjj}_{135}$ through a channel composed by two GHZ states. Measurements in the GHZ-like basis
$\{\ket{\Psi^{\mu}_{\lambda\omega}}\}$ are represented by dotted  and dot-dashed lines, while entanglement between qubits by dashed lines.}
\label{fig2}
\end{figure}

Of course, in order to make comparisons the input state, qubits 1-3-5, is exactly the same as before, Eq. (\ref{input}).
Furthermore, the channel corresponds to two tripartite states, which brings an important difference between the 3-EPR 
and 2-GHZ setups. In the first case qubits 2, 4, and 6 are completely equivalent, the same holding for qubits 7, 8, and 9.
This is not possible in distributing these six qubits between two tripartite states. The only configuration that keeps the original distribution is 
the alternate geometry shown in figure \ref{fig2}. In this situation, qubits 2 and 6 are not equivalent to qubit 4, the same being valid for
qubits 7 and 9 with respect to qubit 8. Analogously to the previous case, we express the density operator associated with the part of the channel involving the qubits $a$, $b$ and $c$ as:
\begin{equation}
\hat{\rho}_{abc}=\sum_{\substack{klmn \\ pq=0,1}} \gamma_{klmnpq}^{(abc)}\ket{klm}\bra{npq}.
\end{equation}

In this case we consider the non-maximally entangled GHZ-like measurement basis for $N=3$ $\{\ket{\Psi^{\mu}_{\lambda\omega}}\}$, whose elements are given 
by (see appendix): $\ket{\Psi_{\lambda\omega}^{\mu}}=\sum_{j=0,1}(-1)^{\mu j}b_{\mu\oplus j}\ket{j,j\oplus\lambda,j\oplus\omega}$, and the coefficient 
$b_{\mu\oplus j}$ defined in the same way as in the previous case. The projector related to the two necessary measurements reads:
\begin{equation}
\hat{P}_{K}= \ket{\Psi^{\mu}_{\lambda\omega}}\bra{\Psi^{\mu}_{\lambda\omega}}_{145}\otimes
\ket{\Psi^{\nu}_{\tau\epsilon}}\bra{\Psi^{\nu}_{\tau\epsilon}}_{236}.
\end{equation}

In principle, there would be $8 \times 8=64$ possible combinations of results (the same number as in the EPR case). However, there are four individual 
outputs that {\it never} occur. These zero-probability results are those with $\omega\ne 0$. Therefore, there are $4 \times 8=2^5=32$ possible results, 
corresponding to 5 bits of classical information and as we will show in the following sections, in several cases there is an additional constraint which reduces to $16$ 
the number of possible results. This means that full measurements in the GHZ-like basis are not necessary, as it happens in the 3-EPR scheme.

After some algebra, we find the local unitary transformations required on the qubits 7, 8, 9:
\begin{equation}
\nonumber
\hat{U}_K= \sum_{klm=0,1} a_{\mu}^{k}a_{\nu}^{l\oplus \tau}\ket{k,l,m}\bra{k \oplus \lambda, l\oplus \tau, m \oplus \lambda},
\end{equation}
which is equivalent to:
\begin{equation}
\hat{U}_{K}=\hat{\sigma}_z^{\mu}\hat{\sigma}_x^{\lambda}\otimes \hat{\sigma}_x^{\tau} \hat{\sigma}_z^{\nu} \otimes \hat{\sigma}_x^{\tau}.
\end{equation}

Thus the fidelity of the teleportation protocol under the 2-GHZ scheme may be calculated. It reads:
\begin{widetext}
\begin{equation}
 \label{F_GHZ}
 F=\sum_{\substack{klmn\mu \\ \nu\tau\epsilon\lambda=0,1}}c_{k'}c_{l'}^*c_{n'}c_{m'}^* a_{\mu\nu}^{klmn} b_{\mu \oplus k '}^* b_{\mu \oplus l '} b_{\nu \oplus l}b_{\nu \oplus k}^*
 \gamma_{k,k\oplus \epsilon,m,l,l\oplus \epsilon,n}^{(268)}\gamma_{k'\oplus\lambda,m'\oplus\lambda,m'\oplus\lambda,l'\oplus\lambda,n'\oplus\lambda,n'\oplus\lambda}^{(479)},
\end{equation}
\end{widetext}
where the primed indexes $j'$ stand for $j\oplus\tau$.

\section{Non-Maximally entangled Channels and measurements}

In this section we assume that the systems in consideration are isolated and that any deviation from ideality comes from imperfections in measurements and in the 
preparation of the channel states.

\subsection{3-EPR Scheme}

The channel of the 3-EPR scheme is composed by three pairs of qubits, which will be assumed to present some systematic deviation from maximal entanglement. We denote the state of each pair composing the channel as $\ket{\psi}=\sum_{j=0,1}\beta_j\ket{jj}$, with 
$\beta_0=\cos\theta$ and $\beta_1=\sin\theta$, which, by inspection of Eq. (\ref{gamma1}) leads to
$\gamma_{klmn}^{(ab)}=\beta_k\beta_m\delta_{kl}\delta_{mn}$. 
By replacing these ingredients in the general expression (\ref{F_EPR}) we get:
\begin{equation}
 F=|c_0|^4+|c_1|^4+128|c_0|^2|c_1|^2 \left(b_0b_1\beta_0\beta_1\right)^3 .
\end{equation}
In order to calculate the average fidelity, the input state can be parametrized as $(c_0,c_1)=(\cos\theta_0,e^{i\varphi}\sin\theta_0)$, with 
$0<\theta_0<\pi/2$ and $0<\varphi<2\pi$. The associated volume element is $dV=\sin\theta_0\cos\theta_0d\theta_0d\varphi$ and the total volume is $V=\pi$. 
The average fidelity of the 3-EPR teleportation scheme under non-maximally entangled channels and measurements takes the form:
\begin{equation}
\label{fepr0}
\langle F_{\rm EPR} \rangle = \frac{2}{3}+ \frac{1}{3}\sin^3(2\theta)\sin^3(2\phi).
\end{equation}
If one assumes that the systematic errors are small, $\theta=\pi/4+\delta \theta$ and  $\phi=\pi/4+\delta \phi$, then, since the first non-vanishing  correction is quadratic 
in the deviations [$\langle F_{\rm EPR} \rangle\approx 1-2(\delta \theta^2+\delta \phi^2)$], the fidelity remains close to $1$. As a numeric example, consider the 
deviations $\delta \theta=\delta \phi= 5^{\rm o}$, which lead to $\langle F_{\rm EPR} \rangle=0.969$. We postpone the analysis of random (non-systematic) errors to 
the final part of this work were several common types of noise will be considered.

\subsection{2-GHZ Scheme}

Following the same approach as in the previous case, the channel is composed by two initially prepared GHZ-like states $\ket{\psi}=\sum_{j=0,1}\beta_j\ket{jjj}$, thus
the $\gamma$ coefficients become $\gamma_{klmnpq}^{(abc)}=\beta_k\beta_n\delta_{kl}\delta_{lm}\delta_{np}\delta_{pq}$. Substituting into the expression for the 
fidelity, Eq. (\ref{F_GHZ}) and after some calculations, we have:
\begin{equation}
% \nonumber
 F=|c_0|^4+|c_1|^4+32|c_0|^2|c_1|^2 \left(b_0b_1\beta_0\beta_1\right)^2.
\end{equation}

In addition, due to the configuration of the channel and 
measurements, the probability of the output corresponding to $\epsilon\ne0$ is null, which reduces the number of possible outputs to 16.

In the same way as before, we calculated the average fidelity of the 2-GHZ teleportation scheme under non-maximally entangled channels and measurements. It reads:
\begin{equation}
\label{fghz0}
\langle F_{\rm GHZ} \rangle = \frac{2}{3}+ \frac{1}{3}\sin^2(2\theta) \sin^2(2\phi),
\end{equation}
which is larger than $\langle F_{\rm EPR} \rangle$, for equal values of $\theta$ and $\phi$. 
For small deviations from ideality,  $\theta=\pi/4+\delta \theta$ and  $\phi=\pi/4+\delta \phi$, we get 
$\langle F_{\rm GHZ} \rangle\approx 1-\frac{4}{3}(\delta \theta^2+\delta \phi^2)$. Taking, as in the previous case, $\delta \theta=\delta \phi= 5^{\rm o}$, we obtain
$\langle F_{\rm GHZ} \rangle=0.979$, which presents a slight improvement with respect to the EPR channel. 
  
\begin{figure}[hpt]
	\centering
	\includegraphics[scale=0.65]{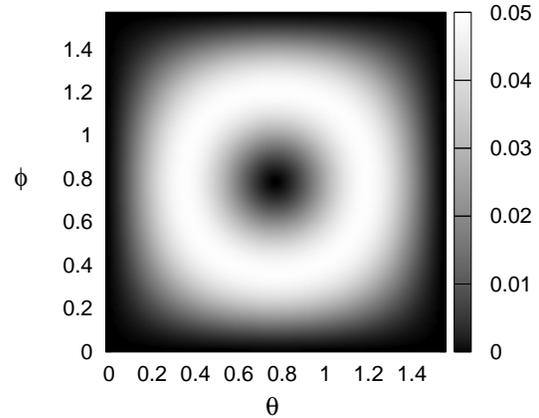}
	\caption{Difference of average fidelities, $\Delta F$ [Eq. (\ref{dif0})], produced by non-maximally entangled channels and measurements, as a function of $\theta$ and $\phi$. The maximal difference 
		$\Delta F=4/81$ is attained for $\sin(2\theta) \sin(2\phi)=2/3$.}
	\label{fig3}
\end{figure}

To have a more comprehensive picture, let us define the difference 
\begin{equation}
\label{dif0}
\Delta F=\langle F_{\rm GHZ} \rangle-\langle F_{\rm EPR} \rangle,
\end{equation}
which is nonnegative, showing that the GHZ channel has a better performance in comparison to the usual EPR channel (see fig. \ref{fig3}). The difference attains its
maximum value whenever channels and measurement bases satisfy $\sin(2\theta) \sin(2\phi)=2/3$. In this situation $\Delta F=4/81\approx 0.049$. In the next section we 
will see that $\Delta F$ may reach even larger values if noise is present.

\section{Weak Noise Regime}
We now address the more realistic situation where, in addition to the systematic errors in the channels and in the swapping operations, random errors may appear. Here 
we will assume that the probabilities of occurrence of these errors in each of the six qubits composing a channel are statistically independent. We denote this probability 
by $p$. In addition, we will limit our analysis to situations where $p$ is small enough so that one can safely disregard the possibility of more than a single error per 
channel. More precisely, the probability that no error occurs in a six-qubit channel is ${\cal P}_0=(1-p)^6$, while the chance that a single error occurs is 
${\cal P}_1=6p(1-p)^5$. We focus on the regime where this probability is much larger than that of two errors per set of six qubits, which is ${\cal P}_2=15p^2(1-p)^4$, 
where we assumed that if the two errors happen in the same qubit, the overall effect is null (valid for bit flip and phase flip). Also we are not taking into account higher 
order events. Therefore a rough estimate for an upper bound for the usefulness of the next results is $p<p_{max}=2/7$ $(\approx 0.29)$.

In the numeric example we provide in the end of this section we consider a probability of error per qubit of $7 \%$ ($p=0.07$) which leads to ${\cal P}_1\approx 0.3$ 
and ${\cal P}_2\approx 0.05$. In this scenario there are only two typical occurrences. Either the channel is free from noise, or a random change happens to a single qubit.
In what follows we address this problem by employing the Kraus-operator formalism to consider that the channels are not completely isolated from the environment.

We will address three common types of noise, namely: (i) bit flip
\begin{equation}
\nonumber
\ket{0} \rightarrow \ket{1}, \hspace{0.25cm} \ket{1} \rightarrow \ket{0},
\end{equation}
corresponding to the Kraus operators
\begin{equation}
\nonumber
\hat{A}_0= \sqrt{1-p}\,\hat{\mathbb{1}}, \hspace{0.4cm} \hat{A}_1= \sqrt{p}\,\hat{\sigma}_x;
\end{equation}
(ii) phase flip 
\begin{equation}
\nonumber
\ket{0} \rightarrow \ket{0}, \hspace{0.25cm} \ket{1} \rightarrow -\ket{1},
\end{equation}
with Kraus operators
\begin{equation}
\nonumber
\hat{A}_0= \sqrt{1-p}\,\hat{ \mathbb{1}}, \hspace{0.4cm} \hat{A}_1= \sqrt{p}\,\hat{\sigma}_z;
\end{equation}
and (iii) depolarizing noise [which corresponds to the individual or combined occurrence of (i) and (ii)]. In this case we have four Kraus operators,
$\hat{A}_0$, $\hat{A}_1$, $\hat{A}_2$, and $\hat{A}_3$, corresponding to
\begin{equation}
\nonumber
\sqrt{1-p}\,\hat{\mathbb{1}}, \hspace{0.4cm} \sqrt{\frac{p}{3}}\,\hat{\sigma}_i,
\end{equation}
respectively, where $\hat{\sigma}_i$ are the Pauli matrices ($i=x, y, z$).

The channels are initially prepared in non-maximally entangled states as in the previous section and the action of noise through the Kraus operators on the state of the 
channels is introduced directly in the $\gamma$ coefficients as we will show in each particular case.
\subsection{Noisy EPR channel}
We are in a position to turn our attention back to the 3-EPR channel subjected to random errors. In all cases
we start by considering that qubit 2 is the one affected by noise. Due to the equivalence between the
three EPR states, the results must be the same for qubits 4 and 6 (see fig. \ref{fig1}). For the EPR channel,
it turns out that the effect of noise is also the same for the distant qubits 7, 8, and 9.
\subsubsection{Bit flip}
We start by considering a bit-flip error in qubit 2, the corresponding $\gamma$ coefficient reads:
\begin{equation}
\nonumber
\gamma_{klmn}^{(27)} = \beta_l \beta_n\Big\{(1-p)\delta_{kl}\delta_{mn}+p\delta_{k,l \oplus 1}\delta_{m,n \oplus 1}\Big\}.
\end{equation}
It is then easy to obtain $\langle F \rangle_{(2)}$ as a function of $p$, $\theta$, and $\phi$ and to show that the result remains unchanged when the error happens in any
of the other five qubits of the channel. The final expression reads:
\begin{equation}
\label{bf-EPR}
\langle F_{\rm EPR}^{\rm B} \rangle = (1-p)\left\{\frac{2}{3}+\frac{1}{3}\sin^3(2\theta) \sin^3(2\phi)\right\},
\end{equation}
where ``B'' stands for bit flip. Therefore, the fidelity is globally affected by this type of noise. A different behavior is observed 
for the other kinds of errors.

\subsubsection{Phase flip}
Let us consider that the qubit 2 is probabilistically subjected to a phase flip, the $\gamma$ coefficient is
\begin{equation}
\nonumber
\gamma_{klmn}^{(27)} = \beta_l\beta_n\delta_{kl}\delta_{mn}\Big\{1-p+p(-1)^{k\oplus m}\Big\}.
% \gamma_{klmn}^{(27)} = \beta_k \beta_{\ell}\Biggl[(1-p)+(-1)^{\ell}p\Biggl].
\end{equation}

Again, the result is the same for all six qubits of the channel, and the final result is
\begin{equation}
\label{pf-EPR}
\langle F_{\rm EPR}^{\rm P} \rangle= \frac{2}{3}+ \frac{1}{3}(1-2p)\sin^3(2\theta)\sin^3(2\phi).
\end{equation}
Note that the classical part of this fidelity is not affected by the amount of noise, which is expected, since classical
bits have no phase whatsoever.
\subsubsection{Depolarizing}
Finally, the $\gamma$ coefficient under depolarizing noise is:
\begin{multline}
\nonumber 
\gamma_{klmn}^{(27)}=\beta_l \beta_n\Big\{ \left(1-p+\frac{p}{3}(-1)^{k\oplus m} \right)\delta_{kl}\delta_{mn}+ \\
+\frac{p}{3}\Big(1+(-1)^{k\oplus m}\Big)\delta_{k,l\oplus 1}\delta_{m,n\oplus 1} \Big\}.
\end{multline}
The total fidelity amounts to:
\begin{equation}
\nonumber
\label{dp-EPR}
\langle F_{\rm EPR}^{\rm D} \rangle=\frac{4}{9}p+\left(1 - \frac{4}{3}p \right)\left(\frac{2}{3}+ \frac{1}{3}\sin^3(2\theta)\sin^3(2\phi)\right).
\end{equation}

In summary, the fidelity is globally compromised when bit-flip noise is present, while only its quantum part is affected by phase flips, as expected. Depolarizing noise 
presents an intermediate result. We recall that, in all situations, the particular channel qubit on which the error occurs is {\it immaterial}.

\subsection{Noisy GHZ channel}
Here we address the same types of noise, this time, acting upon a channel composed by two GHZ states. As we will see, the results may be quite different, both, 
qualitatively and quantitatively. As before, we initially consider that the qubit 2 may suffer random modifications. But, now, due to the distinct geometric distribution of 
entanglement, it is evident that qubit 2 and qubit 4, for instance, are inequivalent.
\subsubsection{Bit flip}
Consider the GHZ-like state of qubits 268 (see figure \ref{fig2}), with a possible flip in qubit 6. In this case the channel coefficient reads
\begin{equation}
\nonumber
\gamma_{klmnpq}^{(268)}=\beta_k \beta_n\delta_{km}\delta_{nq} \Big\{(1-p)\delta_{kl}\delta_{np} + p\delta_{l,k\oplus 1}\delta_{p,n\oplus 1}\Big\}.
\end{equation}
After measurements, classical communication and unitary operations we get a quite remarkable result:
\begin{equation}
\label{!!}
\langle F \rangle_{(6)} = \frac{2}{3}+ \frac{1}{3}\sin^2(2\theta)\sin^2(2\phi),
\end{equation}
which means that qubit 6 is fully protected from bit-flip noise.
In contrast, when the same kind of noise is considered for the other five qubits, one obtains the ordinary result
\begin{equation}
\nonumber
\langle F \rangle_{(j)} = (1-p)\langle F \rangle_{(6)},
\end{equation} 
where $j=2,4,7,8,9$. The final fidelity is given by $[\langle F \rangle_{(6)}+5\langle F \rangle_{(2)}]/6 $, that is:
\begin{equation}
\nonumber
\label{bf-GHZ}
\langle F_{\rm GHZ}^{\rm B}\rangle= \left(1-\frac{5}{6}p\right)\left\{\frac{2}{3}+ \frac{1}{3}\sin^2(2\theta)\sin^2(2\phi)\right\}.
\end{equation} 
\subsubsection{Phase flip}
In the case of phase-flip noise, all six qubits in the channel become equivalent. Particularly, if qubit 2 is subject to phase-flip noise, we have:
\begin{equation}
\nonumber
\gamma_{klmnpq}^{(268)}=\beta_k \beta_n \delta_{kl}\delta_{lm}\delta_{np}\delta_{pq} \Big\{1-p+ p(-1)^{k\oplus n}\Big\}.
\end{equation}
The final result is, in what concerns noise, analogous to that of the 3-EPR scheme:
\begin{equation}
\label{fp-GHZ}
\langle F_{\rm GHZ}^{\rm P} \rangle = \frac{2}{3} + \frac{1}{3}(1-2p)\sin^2(2\theta) \sin^2(2\phi).
\end{equation} 
The difference between $\langle F_{\rm GHZ}^{P} \rangle$ and $\langle F_{\rm EPR}^{P} \rangle $, comes exclusively from systematic errors. 
\subsubsection{Depolarizing}
Since depolarizing is a composition of the two previous noises, qubit 6 presents a different result from the other five. The $\gamma$ coefficient is:
\begin{multline}
\nonumber 
\gamma_{klmnpq}^{(268)}=\beta_{k} \beta_n \delta_{km}\delta_{nq} \Big\{ \left(1-p+\frac{p}{3}(-1)^{l\oplus p} \right)\times \\
\times\delta_{kl}\delta_{np}+\frac{p}{3}\Big(1+(-1)^{l\oplus p}\Big)\delta_{l,k\oplus 1}\delta_{p,n\oplus 1} \Big\}.
\end{multline}
For qubit 6 we obtain:
\begin{equation}
\nonumber
\langle F \rangle_{(6)}= \frac{2}{3} + \frac{1}{3}\left(1-\frac{4}{3}p \right)\sin^2(2\theta) \sin^2(2\phi),
\end{equation} 
while:
\begin{equation}
\nonumber
 \langle F\rangle_{(j)}= \frac{2}{3}\left(1 -\frac{2}{3}p \right) +\frac{1}{3}\left(1-\frac{4}{3}p \right)\sin^2(2\theta) \sin^2(2\phi),
\end{equation} 
with $j=2,4,7,8,9$.
Finally:
\begin{equation}
\nonumber
\label{dp-GHZ}
\langle F_{\rm GHZ}^{\rm D} \rangle= \frac{2}{3}\left(1 -\frac{5}{9}p \right) + \frac{1}{3}\left(1-\frac{4}{3}p \right)\sin^2(2\theta)\sin^2(2\phi).
\end{equation} 
\section{Discussion and Conclusion}
\begin{table}
\caption{\label{tab1}
Overall fidelity delivered by ideal 3-EPR and 2-GHZ schemes, with ideal measurements and noise. }
\begin{ruledtabular}
\begin{tabular}{ccc}
Noise & 3-EPR & 2-GHZ\\
\hline
\\
\quad Bit flip & $1-p $ & $ 1- \frac{5}{6}p $\\
\\
\hline
\\
\quad Phase flip & $ 1- \frac{2}{3}p $ & $ 1- \frac{2}{3}p $ \\
\\
\hline
\\
\quad Depolarizing & $1- \frac{8}{9}p $ & $ 1- \frac{22}{27}p$ \\
\\
\end{tabular}
\end{ruledtabular}
\end{table}

Let us now summarize our results. In the absence of noise and with ideal (maximally entangled) channels and measurements the two kinds of structures (3-EPR and 2-GHZ) lead to a fidelity of 100$\%$. Still, while the 3-EPR scheme requires complete measurements in each run, 
the 2-GHZ scheme demands partial measurements.

As soon as systematic errors appear, in the form of non-maximally entangled channels and measurements, the fidelity obtained with the 2-GHZ setup is consistently higher than that of the 3-EPR setup, see equations (\ref{fepr0}), (\ref{fghz0}), (\ref{dif0}), and fig. \ref{fig3}. These results consider that all involved qubits are fully protected from environmental disturbances. Also, in this case no full measurements in the GHZ basis are required.

The opposite limiting situation is to consider ideal, maximally entangled measurements and channels ($\theta=\phi=\pi/4$), with noise afflicting the latter. These results are shown in table \ref{tab1} where, again, the 2-GHZ scheme presents better performances for bit-flip and depolarizing noises. Particularly remarkable is the fact that for any extent of bit-flip noise, due to the 2-GHZ structure of entanglement distribution and measurements, qubit 6 remain fully protected. For phase flips the two structures lead to the same fidelity.

\begin{figure}
\centering
\includegraphics[scale=0.65]{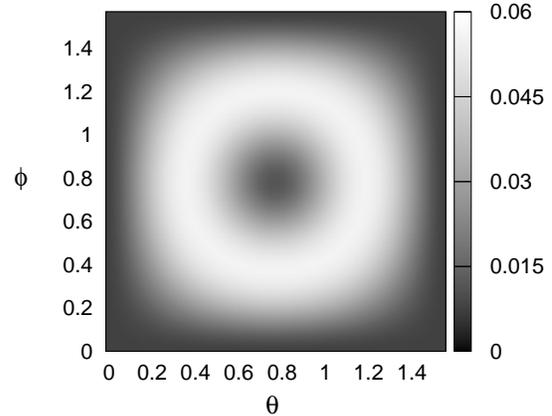}
\caption{Plot of the difference of overall fidelities, $\Delta F^{\rm B}$ [Eq. (\ref{dif1})], produced by 2-GHZ and 3-EPR channels, as a function of $\theta$ and $\phi$ for $p=0.07$. Here, as in fig. \ref{fig3}, black corresponds to $\Delta F^{\rm bf}=0$. The lighter colors present in this plot indicate that the difference between the two setups is more pronounced when noise is present.}
\label{fig4}
\end{figure}

When all these imperfections are considered together, the 2-GHZ scheme is consistently more efficient than the 3-EPR one, as it can be seen by comparing the expressions for average fidelity in each particular case.
The difference between the two schemes is more pronounced for non-ideal entanglement and bit-flip noise (see fig. \ref{fig4}). In this case, it is easy to show that
\begin{equation}
\label{dif1}
\Delta F^{\rm B}=\langle F_{\rm GHZ}^{\rm B} \rangle-\langle F_{\rm EPR}^{\rm B}  \rangle,
\end{equation}
reaches its maximum for:
\begin{equation}
\nonumber
\sin(2\theta^*) \sin(2\phi^*)=\frac{2(1-\frac{5}{6}p)}{3(1-p)}.
\end{equation}
For $p=0.1$ we have $\Delta F^{B}\approx 0.058$.

Our general conclusion is that, although EPR channels are extremely versatile in a wide variety of tasks, channels with more complex entanglement may be more resilient against systematic imperfections and noise in specific tasks.
\begin{acknowledgements}
Financial support from Conselho Nacional de Desenvolvimento Cient\'{\i}fico e Tecnol\'ogico (CNPq) through its program INCT-IQ, Coordena\c{c}\~ao de Aperfei\c{c}oamento de Pessoal de N\'{\i}vel Superior (CAPES), and Funda\c{c}\~ao de Amparo \`a Ci\^encia e Tecnologia do Estado de Pernambuco (FACEPE) is acknowledged.
\end{acknowledgements}

\appendix*
\section{GHZ-Like Basis}
In this appendix we express the general GHZ basis for an arbitrary number of qubits $N$. Any of its elements may be written as:
\begin{table}[b]
\caption{Maximally entangled basis $\{\ket{\Phi_{\lambda}^{\mu}}\}$.}
%\begin{ruledtabular}
\label{tab2}
\begin{tabular}{ccc}
\hline\hline
\quad $\mu$  \quad & \quad $\lambda$ \  \qquad & \quad \ $\ket{\Phi_{\lambda}^{\mu}}$ \qquad\\
\hline
$0$ & $0$ & $\frac{1}{\sqrt{2}}(\ket{00}+ \ket{11}) $ \\
\hline
$0$ & 1 & $\frac{1}{\sqrt{2}}(\ket{01}+ \ket{10}) $ \\
\hline
$1$ & $0$ & $\frac{1}{\sqrt{2}}(\ket{00}- \ket{11}) $ \\
\hline
$1$ & $1$ & $\frac{1}{\sqrt{2}}(\ket{01}- \ket{10}) $ \\
\hline\hline
\end{tabular}
%\end{ruledtabular}
\end{table}
\begin{table}[b]
\caption{Maximally entangled basis $\{\ket{\Phi_{\lambda\omega}^{\mu}}\}$.}
\label{tab3}
\begin{tabular}{cccc}
\hline\hline
\quad $\mu$  \quad & \quad $\lambda$ & \quad $\omega$ \  \qquad & \quad \ $\ket{\Phi_{\lambda\omega}^{\mu}}$ \qquad\\
\hline
$0$ & $0$ & $0$  & $ \frac{1}{\sqrt{2}}(\ket{000}+ \ket{111}) $ \\
\hline
$0$ & $0$ & $1$ & $\frac{1}{\sqrt{2}}(\ket{001}+ \ket{110}) $ \\
\hline
$0$ & $1$ & $0$ & $\frac{1}{\sqrt{2}}(\ket{010}+ \ket{101}) $ \\
\hline
$0$ & $1$ & $1$ & $\frac{1}{\sqrt{2}}(\ket{011}+ \ket{100}) $ \\
\hline
$1$ & $0$ & $0$ & $\frac{1}{\sqrt{2}}(\ket{000}- \ket{111}) $ \\
\hline
$1$ & $0$ & $1$ & $\frac{1}{\sqrt{2}}(\ket{001}- \ket{110}) $ \\
\hline
$1$ & $1$ & $0$ & $\frac{1}{\sqrt{2}}(\ket{010}- \ket{101}) $ \\
\hline
$1$ & $1$ & $1$ & $ \frac{1}{\sqrt{2}}(\ket{011}- \ket{100}) $ \\
\hline\hline
\end{tabular}
\end{table}

\begin{equation}
\ket{\Phi_{\vec{\lambda}}^{\mu}}=\sum_{j=0,1}\frac{(-1)^{\mu j}}{\sqrt{2}}\ket{j}\bigotimes_{k=2}^N\ket{j\oplus\lambda_k},
% \ket{\Psi_{\lambda_2,...,\lambda_N}^{\mu}}=\sum_{j=0}^1\frac{(-1)^{\mu j}}{\sqrt{2}}\ket{j}\bigotimes_{k=2}^N\ket{j\oplus\lambda_k},
\end{equation}

where the parameters that characterize each ket, $\vec{\lambda}=(\lambda_2,...,\lambda_N)$ and $\mu$ may be equal to $0$ or $1$. 
Thus, for instance $\ket{\Phi_{0}^{0}}$ is equal to the element $\ket{\Phi^{+}}$ of the Bell basis and $\ket{\Phi_{00}^{0}}$ corresponds to the standard GHZ state, 
as illustrated in tables \ref{tab2} and \ref{tab3}.

The basis can be generalized even more if we consider non-maximally entangled states:
\begin{equation}
% \ket{\Psi_{\vec{\lambda}}^{\mu}(\phi)}=\sum_{j=0,1}(-1)^{\mu j}b_{\mu\oplus j}\ket{j}\bigotimes_{k=2}^N\ket{j\oplus\lambda_k},
\ket{\Psi_{\vec{\lambda}}^{\mu}}=\sum_{j=0,1}(-1)^{\mu j}b_{\mu\oplus j}\ket{j}\bigotimes_{k=2}^N\ket{j\oplus\lambda_k},
\end{equation}
where $b_{\mu\oplus j}$ controls the degree of entanglement of the basis, which is maximal when $b_{\mu\oplus j}=1/\sqrt{2}$. Therefore it is natural to parametrize
the coefficients as $b_0=\cos\phi$ and $b_1=\sin\phi$ ($0<\phi<\pi/2$), in order to ensure normalization.
It may be easily shown that the elements of the GHZ-like basis satisfy the orthonormality relation:
\begin{equation}
\bra{\Psi_{\vec{\lambda}}^{\mu}}\Psi_{\vec{\tau}}^{\nu}\rangle=\delta_{\vec{\lambda},\vec{\tau}}\delta_{\mu,\nu}.
\end{equation}

\end{document}